\documentclass[a4paper,twoside]{article}
\newcommand{\affil}[1]{$^{\rm #1}$}

\def \low{SKA-low}

\def \AEff{A$_{\rm eff}$}
\global\long\def\AEmaths{A_{\rm eff}}

\def \A_T{$\frac{A_{\rm e}}{T_{\rm sys}}$}
\def \NE{N$_{\rm e}$}
\global\long\def\NEmaths{N_{\rm e}}
\def \Dst{D$_{\rm st}$}
\global\long\def\Dstmaths{D_{\rm st}}

\def \deg{$^{\circ}$}
\usepackage[authoryear]{natbib}
\bibpunct{(}{)}{;}{a}{}{,}
\usepackage{graphicx}

\usepackage[gen]{eurosym} 

\usepackage{subfig}
\usepackage{url}
\makeatother

\date{} 
\title{\large\bf\flushleft Square Kilometre Array station configuration using two-stage beamforming}
\author{\parbox{\textwidth}{\flushleft
\vspace{-0.5cm}
{\it Aziz Jiwani\affil{A, D}, Tim Colegate\affil{A}, Nima Razavi-Ghods\affil{B}, Peter J Hall\affil{A}, Shantanu Padhi\affil{A} and Jan Geralt bij de Vaate\affil{A, C}}\\
\vspace{0.4cm}
{\small \affil{A}\, International Centre for Radio Astronomy Research, Curtin University, GPO Box U1987, Perth, WA 6845, Australia}\\
{\small \affil{B}\,Cavendish Laboratory,  University of Cambridge, JJ Thomson Avenue, Cambridge CB3 0HE, UK }\\
{\small \affil{C}\,ASTRON, P.O. Box 2, 7990 AA Dwingeloo, The Netherlands}\\
{\small \affil{D}\, Email: aziz.jiwani@icrar.org}}}
\begin{document}
\twocolumn[
\begin{changemargin}{.8cm}{.5cm}
\begin{minipage}{.9\textwidth}
\vspace{-1cm}
\maketitle
%
%
\small{\bf Abstract:} The lowest frequency band (70--450~MHz) of the Square Kilometre Array will consist of sparse aperture arrays grouped into geographically-localised patches, or stations. Signals from thousands of antennas in each station will be beamformed to produce station beams which form the inputs for the central correlator. Two-stage beamforming within stations can reduce \low\ signal processing load and costs, but has not been previously explored for the irregular station layouts now favoured in radio astronomy arrays. This paper illustrates the effects of  two-stage beamforming on sidelobes and effective area, for two representative station layouts (regular and irregular gridded tile on an irregular station). The performance is compared with a single-stage, irregular station. The inner sidelobe levels do not change significantly between layouts, but the more distant sidelobes are affected by the tile layouts; regular tile creates diffuse, but regular, grating lobes. With very sparse arrays, the station effective area is similar between layouts. At lower frequencies, the regular tile significantly reduces effective area, hence sensitivity. The effective area is highest for a two-stage irregular station, but it requires a larger station extent than the other two layouts. Although there are cost benefits for stations with two-stage beamforming, we conclude that more accurate station modelling, and \low{} configuration specifications, are required before design finalisation.

\medskip{\bf Keywords:} aperture arrays --- beamforming --- SKA --- radio astronomy

\medskip
\medskip
\end{minipage}
\end{changemargin}
]
\small

\section{Introduction}
The Square Kilometre Array (SKA) is being developed as a radio telescope with a combination of unprecedented sensitivity, resolution and field of view (FoV) over a 70 MHz to $>$10~GHz frequency range~\citep{dewdney2009}. The key scientific objectives of this instrument include mapping the distribution of neutral hydrogen gas in the Universe, studying the origin of magnetic fields and measuring the dynamics of pulsars in the Galaxy. To realize these objectives, a very large collecting area is needed (typically one square kilometre) using new antenna technology that can be mass produced with minimum cost. 

The SKA design consists of three frequency bands, the lowest of which, termed \low\ \citep{memo125}, has been selected to be built in Australia and New Zealand\footnote{\url{http://www.skatelescope.org/news/dual-site-agreed-square-kilometre-array-telescope}}. \low{} will use aperture phased arrays of antennas operating in the 70--450 MHz range and extending across a distance of several hundred kilometres. The wideband antennas convert cosmic radio waves from physical processes in the universe to RF signals, which are digitised, cross-correlated and averaged. Post-correlation aperture synthesis is used to reconstruct the brightness distribution, to achieve high-dynamic range radio pictures of the sky~\citep{HalSch08, memo125}. 

\begin{figure}
	\begin{center}
	\includegraphics[width = \columnwidth]{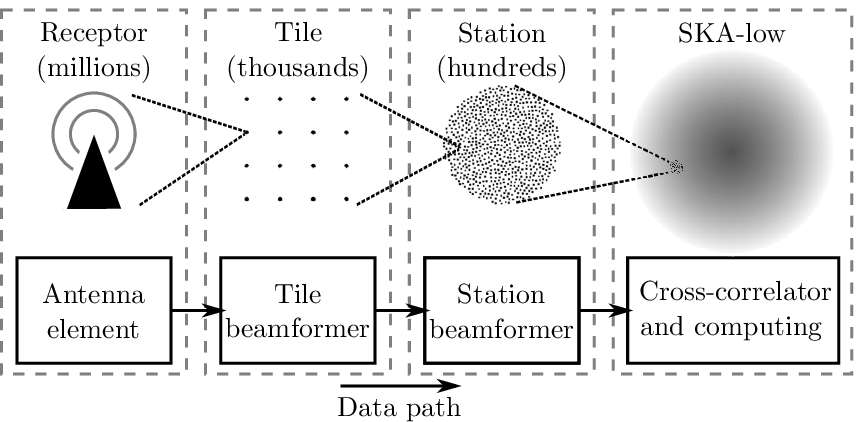}
	\caption{\low\ system schematic.}\label{SKA-sch}
	\end{center}
\end{figure}

Cross-correlating the several million \low\ antenna elements signals to create these radio pictures would require an extremely large signal processing system. Instead, the signal processing load is reduced by grouping geographically-adjacent antenna elements into digitally-beamformed stations; this limits the number of signals to be transported and cross-correlated. Additional cost efficiencies can be gained by grouping individual antenna elements into beamformed tiles prior to the station beamformer~\citep{FauAle10}. Figure~\ref{SKA-sch}~shows a schematic of this two-stage beamforming approach for \low.  The cost efficiencies  arise because the tile beamformer limits the number of independent signals (analogue or digital) early in the signal path. A first-order approximation of this effect is described in Appendix~\ref{app_a}.

The actual reduction in station hardware cost due to two-stage beamforming depends on the technologies used. When analogue first-stage and digital second-stage beamforming is implemented in place of two-stage digital beamforming, the cost reduction is larger. This is because the analogue first-stage beamforming reduces, by an order of magnitude, the digitisation, intra-station signal transmission and station beamformer costs. For example, an analysis of SKA-low station hardware costs shows a factor between 1.5 and 5 reduction in station hardware cost compared to all-digital beamforming \citep{ColHal12}, across various architectural implementations. With cost estimates in \citet{FauVaa11-Deployment} of \euro1.8--2.8\,M per station, this is a potentially significant cost reduction.

Present day telescopes have various approaches to tile or station beamforming. The high band antenna (HBA) of the Low Frequency Array (LOFAR) telescope~\citep{de2009lofar} uses a two-stage beamforming station, but both stages are performed on regular gridded antenna elements, making it an example of a regular tile in a regular station.  The  low band antenna (LBA)  of the same telescope uses a single stage of beamforming on antenna elements in an irregular station layout. The Long Wavelength Array Station 1 (LWA1)  telescope~\citep{EllTay12} is an irregular array with a single-stage of beamforming, and is the first station of  an envisioned larger correlation array. Instead of a station approach, the Murchison Widefield Array (MWA) telescope~\citep{lonsdale2009murchison, TinGoe12} feeds a many-input (large-N) correlator with beamformed 16-element regular tiles.

None of these telescopes use beamformed tiles within an irregular station layout. However, stations composed of an irregular array of elements are under active consideration for the SKA \citep{BijLer11}. Because these stations will have several orders of magnitude more elements per station than these other telescopes, two-stage beamforming is an attractive option to reduce cost. Although antenna elements arranged into beamformed tiles can change key performance characteristics of the station, the effects of two-stage beamforming on the beam pattern and effective area of an irregular station have not been considered previously in the SKA context. This paper examines the implications of placing beamformed tiles within an irregular station grid layout, and compares it with a single-stage irregular station. The aim in doing so is to assess the magnitude of performance degradation which accompanies the potentially lower-cost two-stage layouts. Even though the intra-station layouts considered in this paper are not optimised for performance, they illustrate the distinguishing features of each layout and provide a reference point for future studies of SKA station layout and system design.

The paper is arranged as follows: Section~\ref{sec:background} sets out the intra-station geometry and the analysis method used. Results are presented in Section~\ref{sec:results} and discussed in Section~\ref{sec:discussion}.

\section{Theoretical Background}\label{sec:background}
Aperture arrays (AAs) must be exquisitely well calibrated to meet the high imaging dynamic range requirements for the SKA~\citep{dewdney2009}; one facet of calibration is correcting for a non-ideal synthesized beams, formed from the correlation of signals from AA stations. To do this, the instrumental response needs to be accurately characterized via an astronomical calibration procedure~\citep{wijnholds2010calibration}. 

The difficulty in calibrating the station beam depends on the station size and the nature of its sidelobes~\citep{WijBre11}, and is influenced by the element configuration; a regular array produces strong, predictable sidelobes while an irregular array produces weaker, more diffuse sidelobes. An irregular array smears out in spatial extent the sidelobes that are present at frequencies where a regular array becomes sparse. These sidelobes, or grating lobes in sparse regular arrays, cause strong frequency and direction dependent variations in the main lobe gain~\citep{CapWij06}. 

For this analysis, the effects of mutual coupling are ignored as they tend to average out for large irregular arrays (or stations)~\citep{gonzalez2011non}. There will be non-trivial mutual coupling effects on a tile level but from a station perspective, these effects will not fundamentally change features such as sidelobe levels and effective area, which we analyse for the two-stage beamforming approach. Future work will incorporate the second order effects of mutual coupling and simulated (or measured) element patterns. 

\begin{table*}
	\begin{center}
	\caption{Station specifications}\label{sim_spec}
		\begin{tabular}{p{8cm} p{2cm} p{2cm} p{2cm}}
		\hline & single-stage & reg-irreg & irreg-irreg \\ \hline
		3dB beamwidth of element & 70$^{\circ}$ & 70$^{\circ}$ & 70$^{\circ}$ \\
		Maximum element footprint & 1.2 m & 1.2 m & 1.2 m \\
		Minimum separation between elements (centre to centre) & 1.3 m & 1.3 m & 1.3 m \\
		Number of elements per tile & NA & 16 & 16 \\
		Tile breadth (centre to centre) & NA & 3.9 m & 6.3 m \\
		Minimum separation between tiles (centre to centre) & NA & 5.5 m & 6.8 m \\
		Number of tiles in a station & NA & 699 & 701 \\
		Station diameter \Dst{} (centre to centre) & 189 m & 189 m & 248 m \\
		Total number of elements in station \NE{} & 11 200 & 11 184 & 11 216 \\
		Number of dual polarised signals paths & 11 200 & 699 & 701 \\
		\hline
		\end{tabular}
	\medskip\\
	\end{center}
\end{table*}

\begin{figure}
	\begin{center}
	\includegraphics[width = \columnwidth]{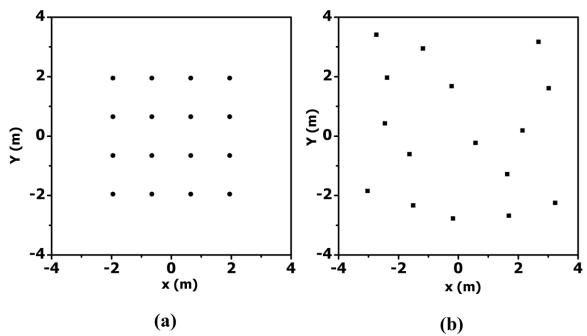}
	\caption{Antenna element layout for (a) regular gridded tile and (b) irregular gridded tile}\label{fig:gridded_tile}
	\end{center}
\end{figure}

\begin{figure}
	\begin{center}
	\includegraphics[width = \columnwidth]{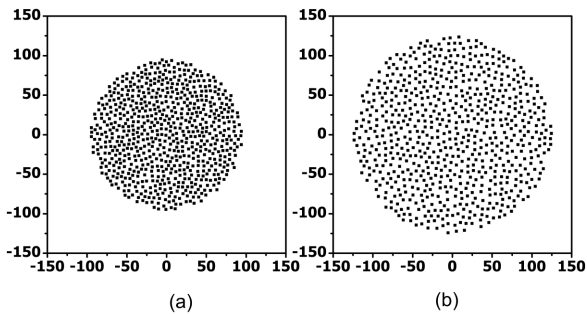}
	\caption{Tile layout for (a) station with regular gridded tile and (b) station with irregular gridded tile}\label{fig:gridded_station}
	\end{center}
\end{figure}

Three representative configurations for station layout are considered in this paper; (i) single-stage irregular array (single-irregular), (ii) regular gridded tile in an irregular array (regular-irregular) and (iii) irregular gridded tile in an irregular array (irregular-irregular). The irregular-irregular configuration uses the same randomised tile layout for all tiles in the station. Both two-stage stations use a single tile layout to enable easily replicated beamforming hardware with a fixed number of inputs. This is likely to reduce the beamformer hardware cost. We note that future work, based in part on the present study, could involve “non-tile” intra-station configurations, wherein signals from free-form antenna groups are beamformed.

Specifications for the representative configurations are given in Table I and the layouts for the two-stage stations are shown in Figure~\ref{fig:gridded_tile} and Figure~\ref{fig:gridded_station}. For comparison, the current description specifies 11~200 elements within a 180~m diameter \low\ station~\citep{Dewbij10}. We note that emerging thinking about the \low\ configuration tends to now favour a somewhat larger number of smaller stations~(e.g.~\cite{mellema2012reionization}).

The number of elements, station diameter and element layout within the station are key parameters in determining station effective area and field of view. We choose to maintain a fixed number of elements in order to characterise the station effective area in the sparse regime of operation where the station sensitivity depends directly on the number of elements. The number of elements per station (\NE) is held approximately constant and a 1.3 m minimum inter-element spacing ($\lambda$/2 at 115~MHz) is enforced. For the two-stage layouts, each tile consists of 16 elements. Due to the randomisation, the irregular tile is necessarily larger than the regular tile; hence a larger station diameter is required for the irregular-irregular layout. An alternative approach is to control station diameter, and vary the number of elements per station. While this approach ensures a similar FoV between station layouts, station costs are not directly comparable because the number of signal chains vary. Furthermore, such an approach for the irregular-irregular layout would result in fewer elements, hence the station high-frequency sensitivity would suffer.  Holding \NE\ constant ensures that both two-stage beamforming stations have the same number of digitised signal chains. 

The simulations are carried out in two stages, using Xarray\footnote{\url{http://sites.google.com/site/xarraytool/}}, a tool developed for computing radiation patterns together with effective area and other aperture array parameters. We generate the tile beam pattern using $\cos{\theta}^{3.47}$ as a single element pattern which approximates one of the moderately directive candidate \low\ antenna radiation pattern~\citep{Eloy_2012_SKALA}. The station beam pattern uses this tile beam as the sub-array pattern for the two-stage beamforming. For single-stage beamforming we simply use the single element pattern in generating the station beam. The beams are calculated with an antenna radiation efficiency of 90$\%$, which is the minimum requirement for the \low\ station~\citep{Dewbij10}.

\section{Results}\label{sec:results}

Conceptually, the SKA-low  stations replace the function of dishes in a radio telescope array; each station beam is an input to the correlator. As for dishes, two important metrics describing telescope performance are the station beam sidelobe levels and station effective area (\AEff{}). 

Unlike dishes, the effective area of a station is strongly dependent on frequency. This occurs because the aperture is not fully-sampled (or not a dense, highly coupled array). Effective area also changes as a function of zenith angle due to the element pattern and geometric effects.  Figure~\ref{fig:aeff} shows \AEff\ of the three station layouts as a function of frequency, for a zenith angle of 0\deg{}.

The sidelobe levels are evident in the station beam pattern. The broadside station beam pattern, at 70~MHz and 300~MHz, for all three cases is shown  in the u-v plane in Figure~\ref{fig:station-beam} and Figure~\ref{fig:station-beam:300}. The same beam at the v = 0 plane for the two frequencies is plotted in Figure~\ref{fig:VO:station-beam} and Figure~\ref{fig:VO:station-beam:300}. The u-v coordinates (equivalent to \textit{l, m} coordinates) are given by u = $\sin{\theta}\cos{\phi}$ and v = $\sin{\theta}\sin{\phi}$ where $\theta$ and $\phi$ are zenith angle and azimuth angle respectively.

Statistics on inter-element spacing can provide further insight into the performance of stations of equal diameter (the single-stage irregular and regular--irregular layouts). The minimum separation between elements in the regular--irregular layout is defined by the regular tile layout: 1.3\,m for every element. The single-stage irregular station has a range of minimum inter-element spacings, defined by physical constraints and randomisation of the elements within these constraints. The 1.3\,m minimum spacing between elements is imposed by the element's footprint, while the 189\,m diameter station boundary limits how far apart the elements can be spaced.  Figure~\ref{fig:1stageStats} shows a frequency count of the minimum distance from each element to all other elements. The mean of these minimum inter-element spacings is 1.38\,m and the maximum is 2.21\,m. For comparison, the minimum inter-element spacing of 11\,200 elements within a 189\,m diameter single-stage regular gridded station (not shown) is approximately 1.6\,m.

\section{Discussion}\label{sec:discussion}

\subsection{Station effective area}

\begin{figure}
	\begin{center}
	\includegraphics[width = \columnwidth]{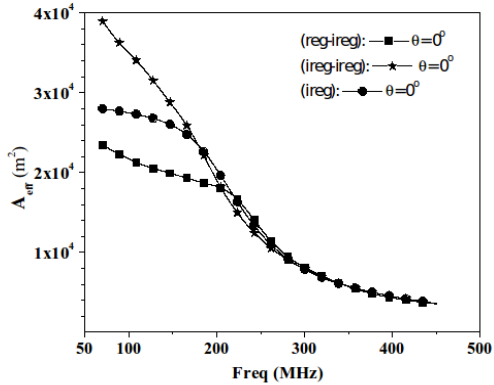}
	\caption{Station zenith \AEff{} as a function of frequency.}\label{fig:aeff}
	\end{center}
\end{figure}

The station effective area \AEff{} is an important metric for \low because, for a constant number of stations, telescope sensitivity is linearly proportional to \AEff{}. A related factor is whether the array is `dense' or `sparse' at the frequency in question. Both \AEff{} and the transition from dense and sparse varies with frequency and the inter-element spacing.

There is no single definition for when an array is dense or sparse. The broad definition used in this paper is that an array is dense when the aperture is fully sampled with $\lambda$/2 or closer element packing, and inter-element mutual coupling is significant.  Effective area is then approximately equal to the physical (geometric) area of the station: 
\begin{equation}
\AEmaths\approx\frac{\pi}{4}\eta\Dstmaths^{2},\label{eq:AeffDense}
\end{equation}
where  $\eta$ is the antenna element radiation efficiency. When the array is sparse, the effective area of each isolated element contributes to the array effective area, such that
\begin{equation}
\AEmaths=\NEmaths\frac{\lambda^{2}}{4\pi}\eta\mathcal{D},
\end{equation}where $\mathcal{D}$ is the directivity of an isolated antenna element~\citep{Bal05}. Wideband aperture arrays have a dense to sparse transition region, which occurs over an inter-element spacing of 0.5-1.5$\lambda$ for dipole-type antennas~\citep{BraCap06}, and typically greater than 2$\lambda$ for more directive antennas~\citep{Rog08-DAM70}.

\begin{figure}
	\begin{center}
	\includegraphics[width = \columnwidth]{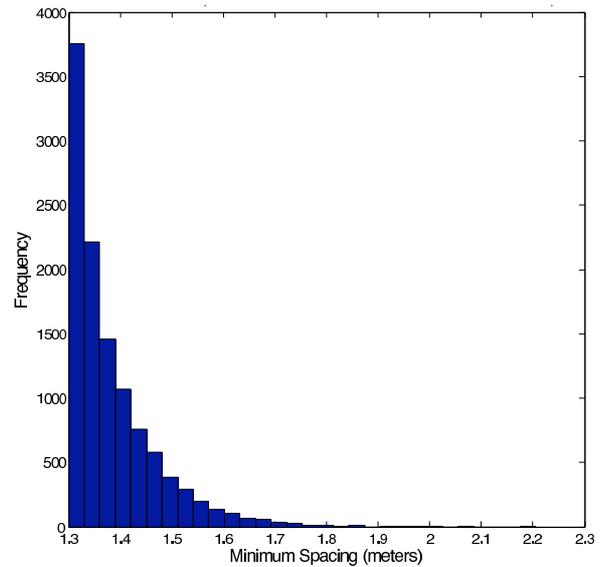}
	\caption{Minimum inter-element spacing for individual elements in the single-stage irregular layout. }\label{fig:1stageStats}
	\end{center}
\end{figure}

As Figure~\ref{fig:aeff} shows, \AEff\ for all layouts is similar at the higher frequencies, where the elements are in the sparse regime and \AEff{}$\propto\lambda^{2}$.  The dense-sparse transition is evident at lower frequencies. Here, effective area is no longer proportional to $\lambda^2$ as mutual coupling effects become active. 

\begin{figure}
	\begin{centering}
	\subfloat[\label{fig:1stage-irreg}]{\centering{}\includegraphics[width=\columnwidth]{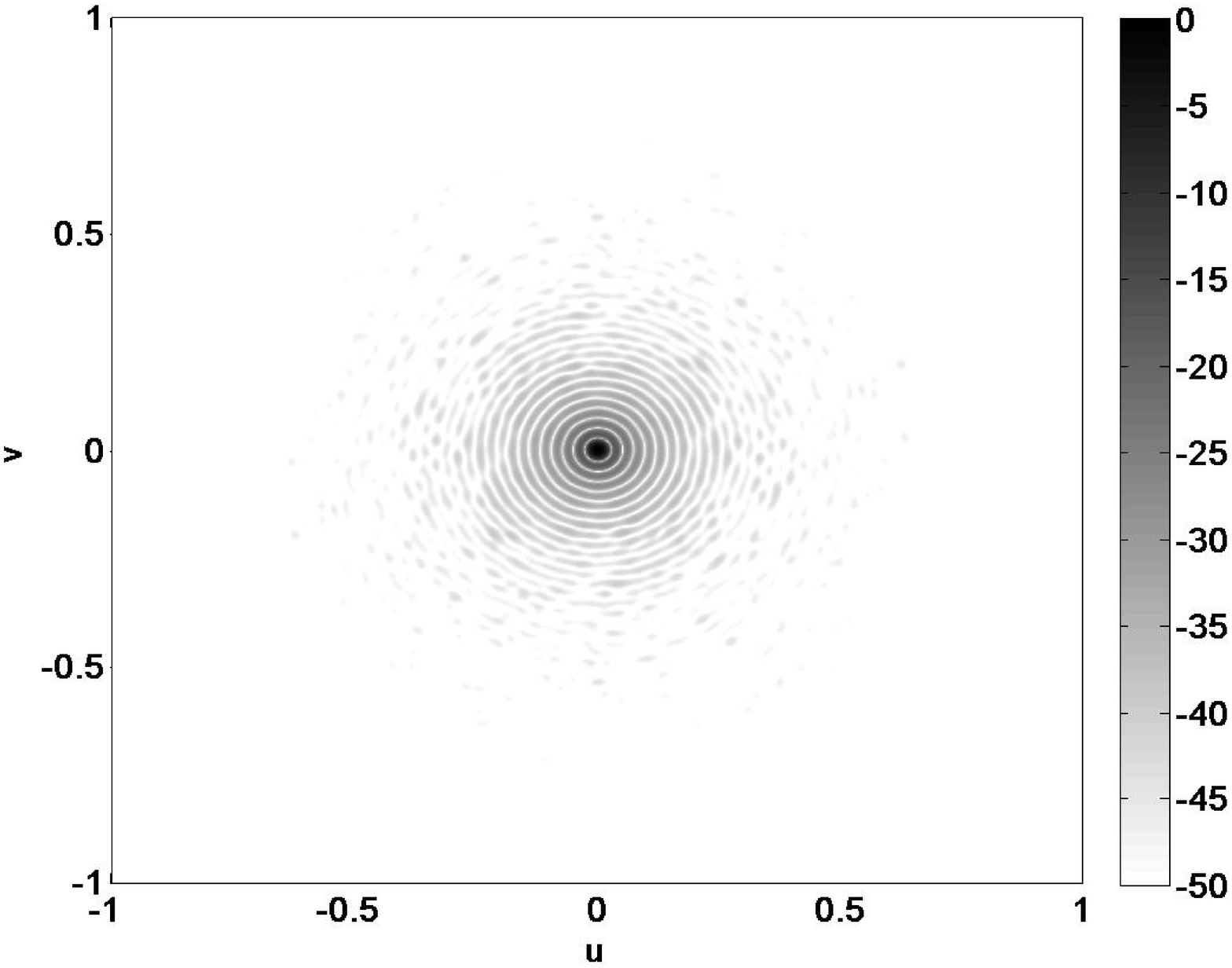}}
	\par\end{centering}
	
	\begin{centering}
	\subfloat[\label{fig:2stage-reg-irreg}]{\centering{}\includegraphics[width=\columnwidth]{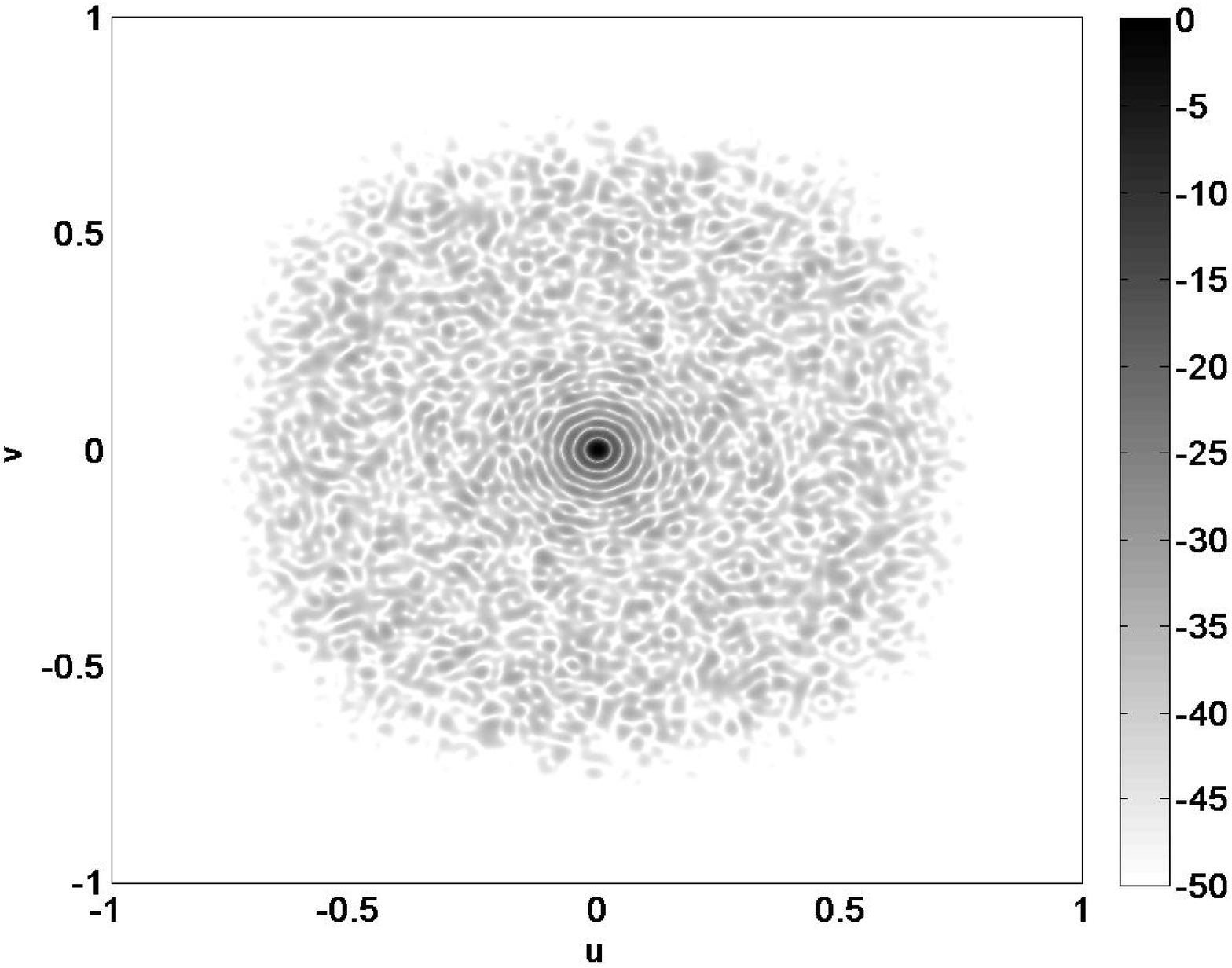}}
	\par\end{centering}
	
	\begin{centering}
	\subfloat[\label{fig:2stage-irreg-irreg}]{\centering{}\includegraphics[width=\columnwidth]{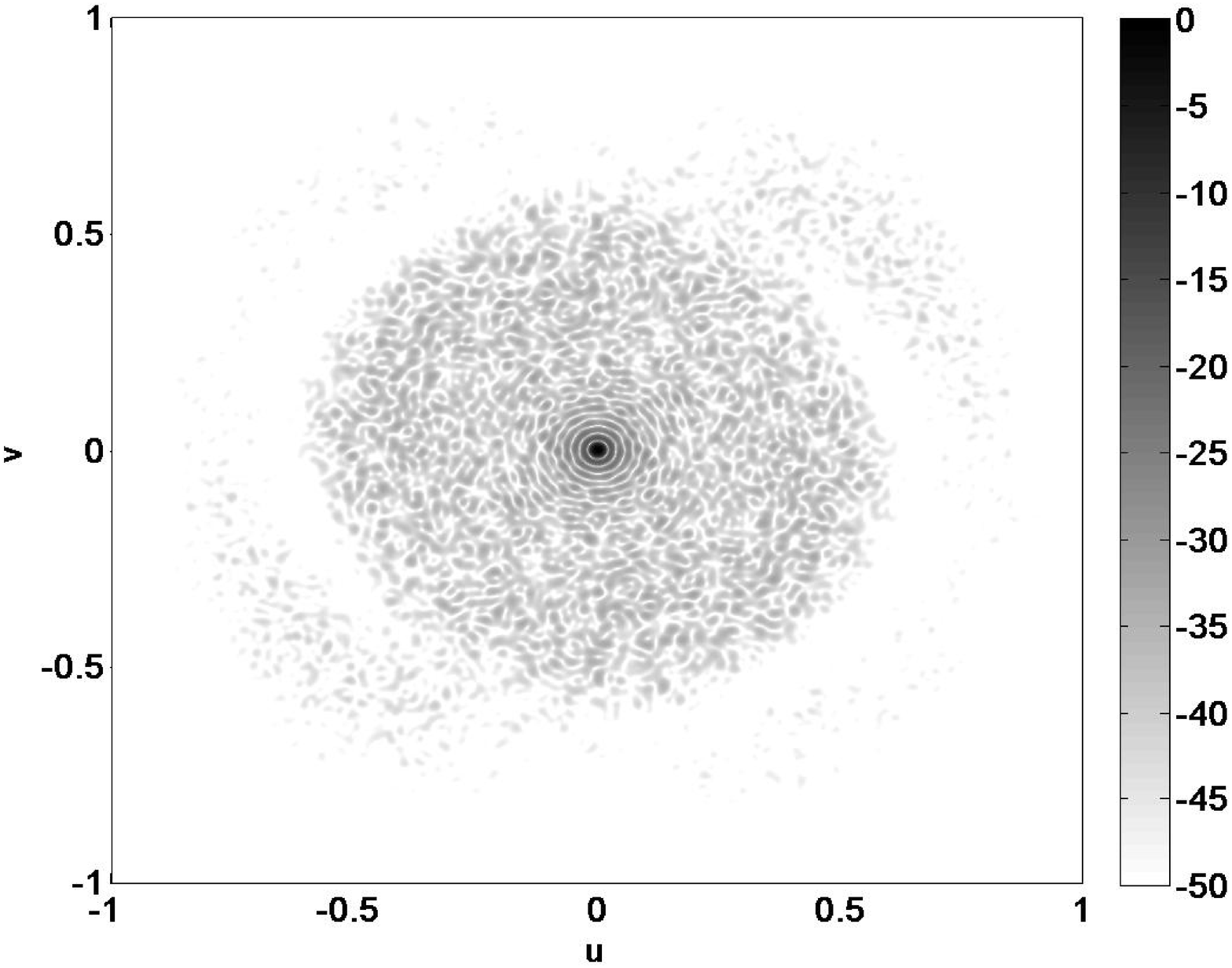}}
\par\end{centering}

	\caption{Broadside station beam at 70~MHz in u-v plane where u = $\sin{\theta}\cos{\phi}$ and v = $\sin{\theta}\sin{\phi}$ ($\theta$ is the zenith angle and $\phi$ is the azimuth angle) for (a) single-stage irregular station layout, (b) two-stage regular-irregular station layout  and (c) two-stage irregular-irregular station layout. 
	\label{fig:station-beam}
}
\end{figure}

\begin{figure}
	\begin{centering}
	\subfloat[\label{fig:VO:1stage-irreg}]{\centering{}\includegraphics[width=\columnwidth]{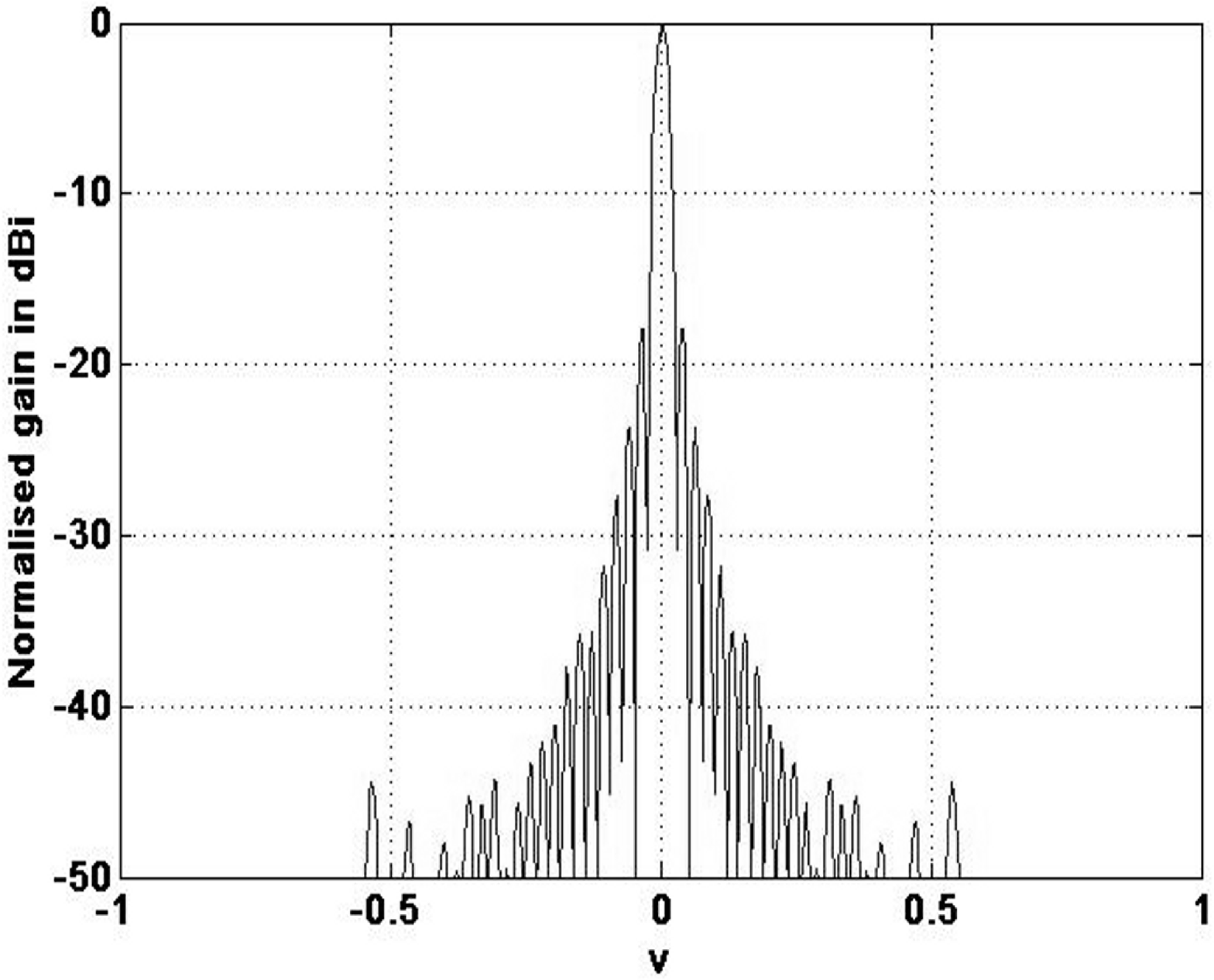}}
	\par\end{centering}
	
	\begin{centering}
	\subfloat[\label{fig:VO:2stage-reg-irreg}]{\centering{}\includegraphics[width=\columnwidth]{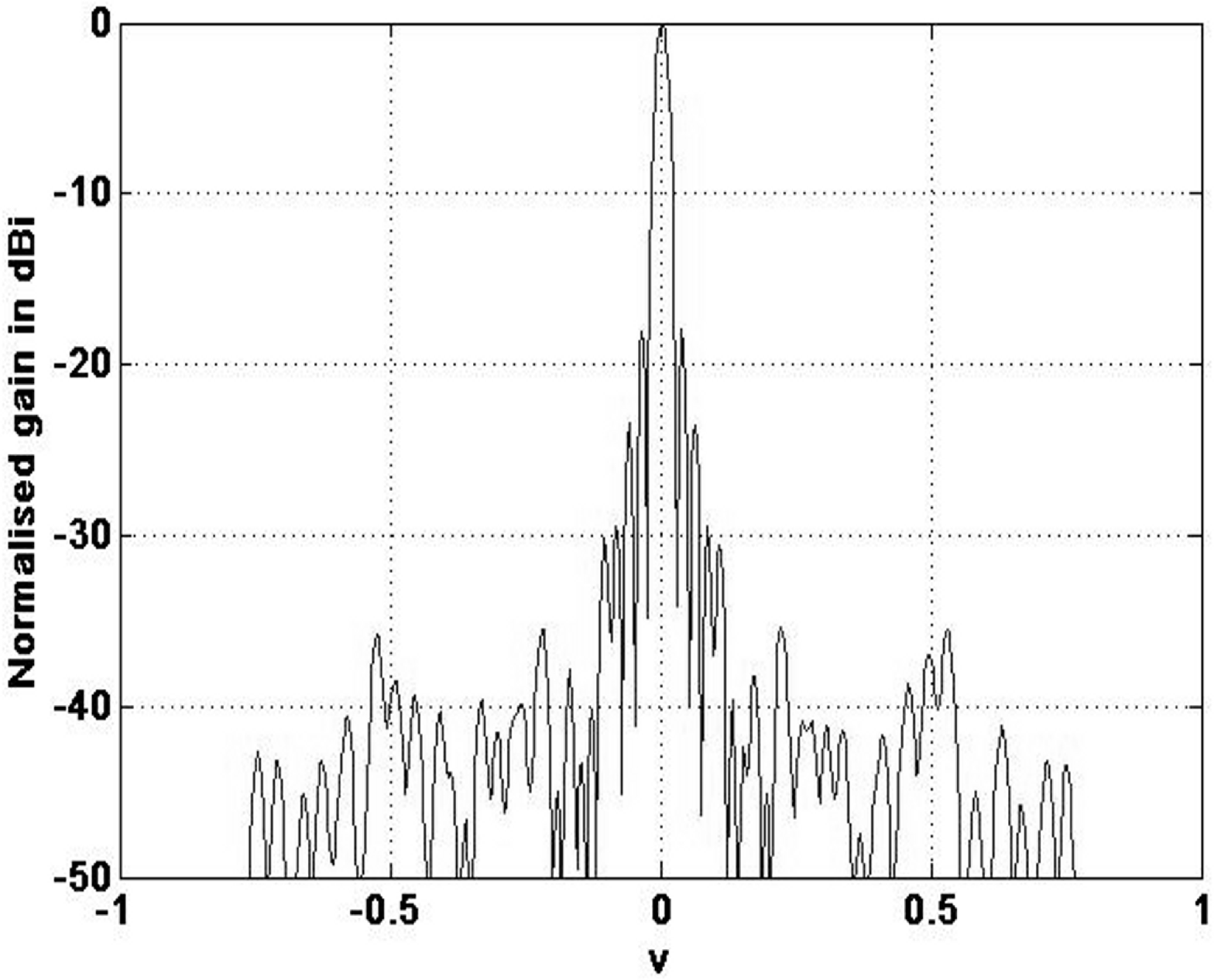}}
	\par\end{centering}
	
	\begin{centering}
	\subfloat[\label{fig:VO:2stage-irreg-irreg}]{\centering{}\includegraphics[width=\columnwidth]{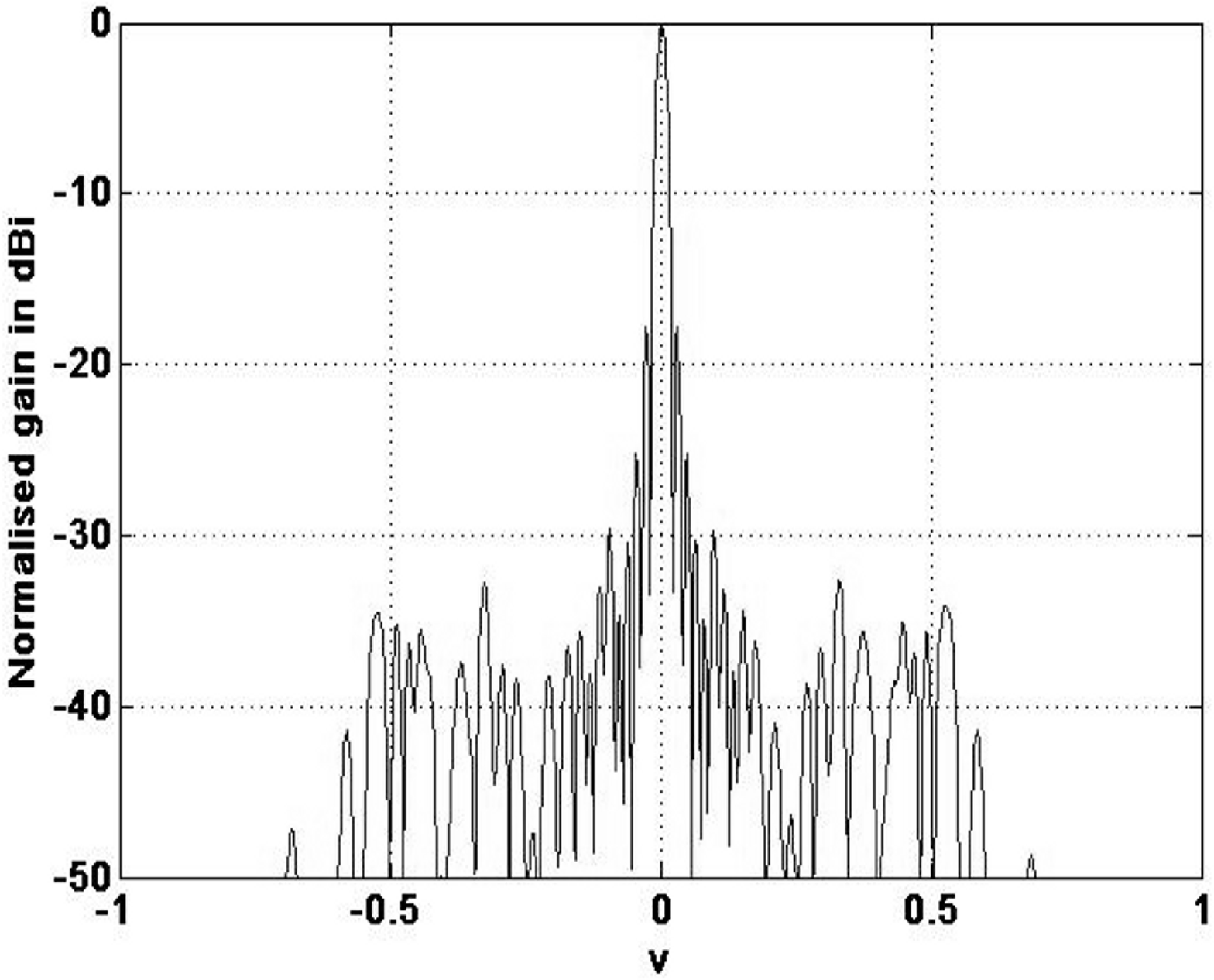}}
\par\end{centering}

	\caption{Broadside station beam at 70~MHz in v=0 plane for (a) single-stage irregular station layout, (b) two-stage regular-irregular station layout  and (c) two-stage irregular-irregular station layout. 
	\label{fig:VO:station-beam}
}
\end{figure}

\begin{figure}
	\begin{centering}
	\subfloat[\label{fig:1stage-irreg:300}]{\centering{}\includegraphics[width=\columnwidth]{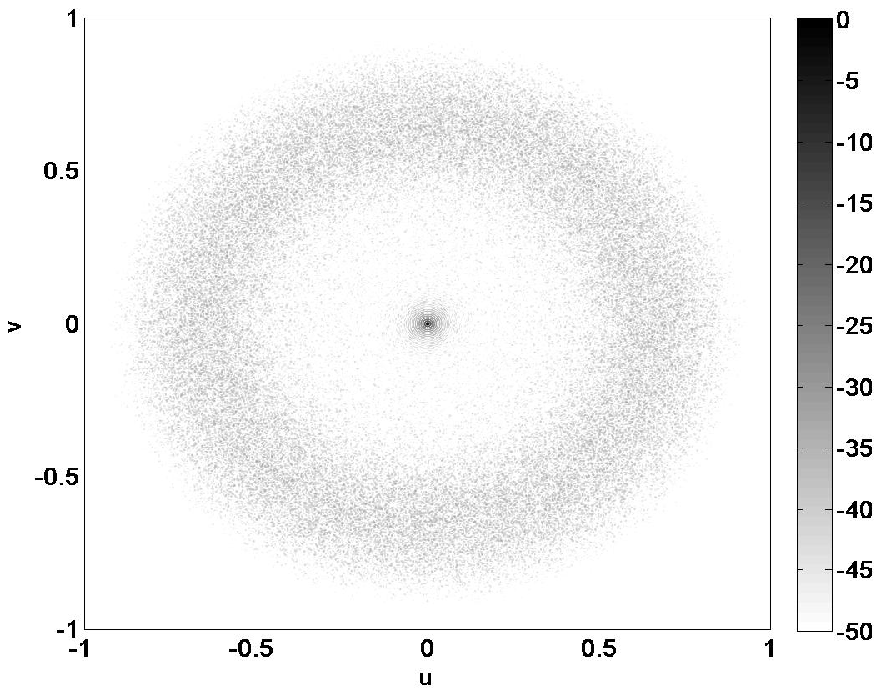}}
	\par\end{centering}
	
	\begin{centering}
	\subfloat[\label{fig:2stage-reg-irreg:300}]{\centering{}\includegraphics[width=\columnwidth]{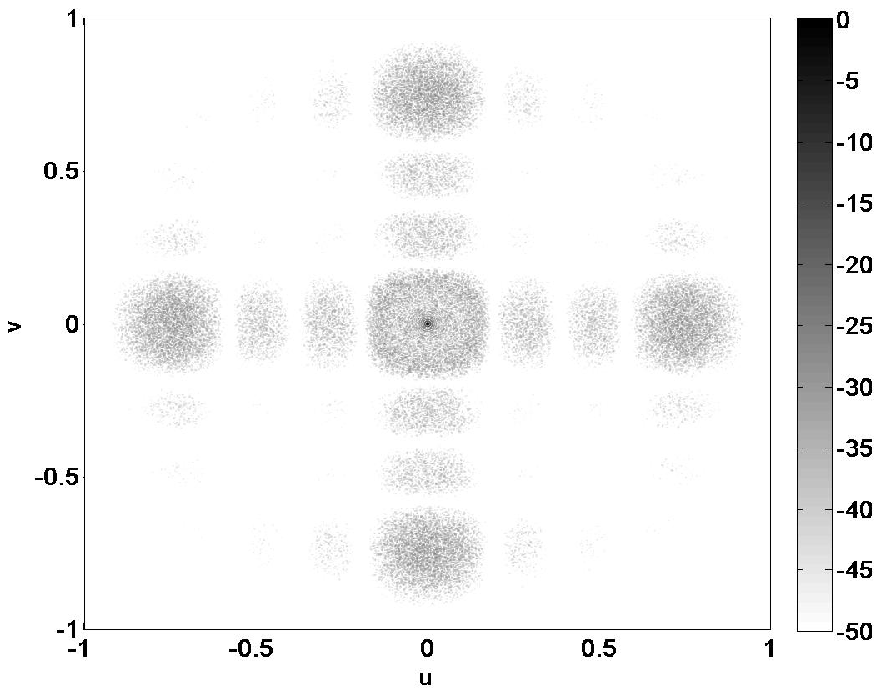}}
	\par\end{centering}
	
	\begin{centering}
	\subfloat[\label{fig:2stage-irreg-irreg:300}]{\centering{}\includegraphics[width=\columnwidth]{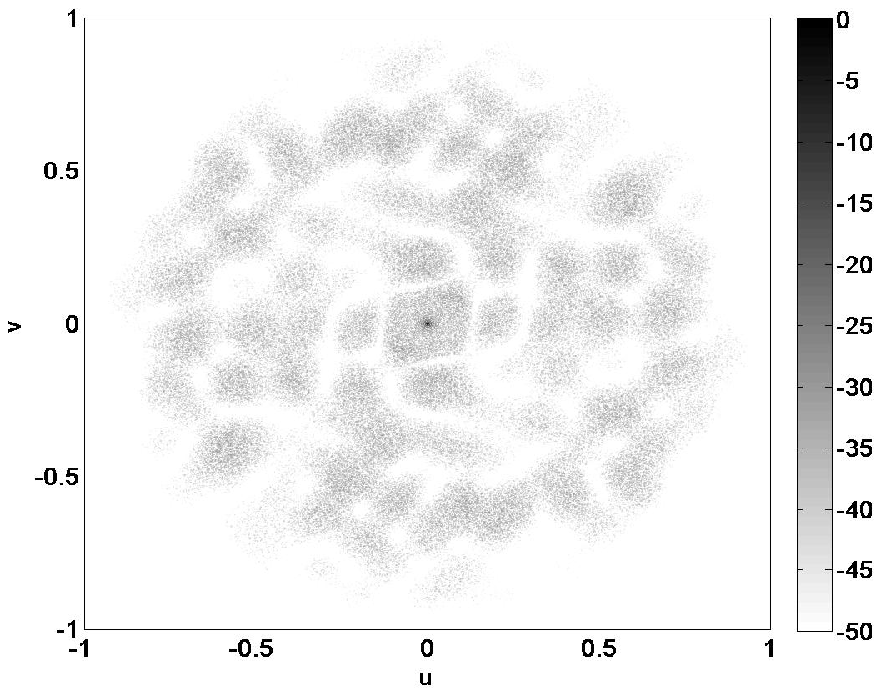}}
\par\end{centering}

	\caption{Broadside station beam at 300~MHz in u-v plane where u = $\sin{\theta}\cos{\phi}$ and v = $\sin{\theta}\sin{\phi}$ ($\theta$ is the zenith angle and $\phi$ is the azimuth angle) for (a) single-stage irregular station layout, (b) two-stage regular-irregular station layout  and (c) two-stage irregular-irregular station layout. 
	\label{fig:station-beam:300}
}
\end{figure}

\begin{figure}
	\begin{centering}
	\subfloat[\label{fig:VO:1stage-irreg:300}]{\centering{}\includegraphics[width=\columnwidth]{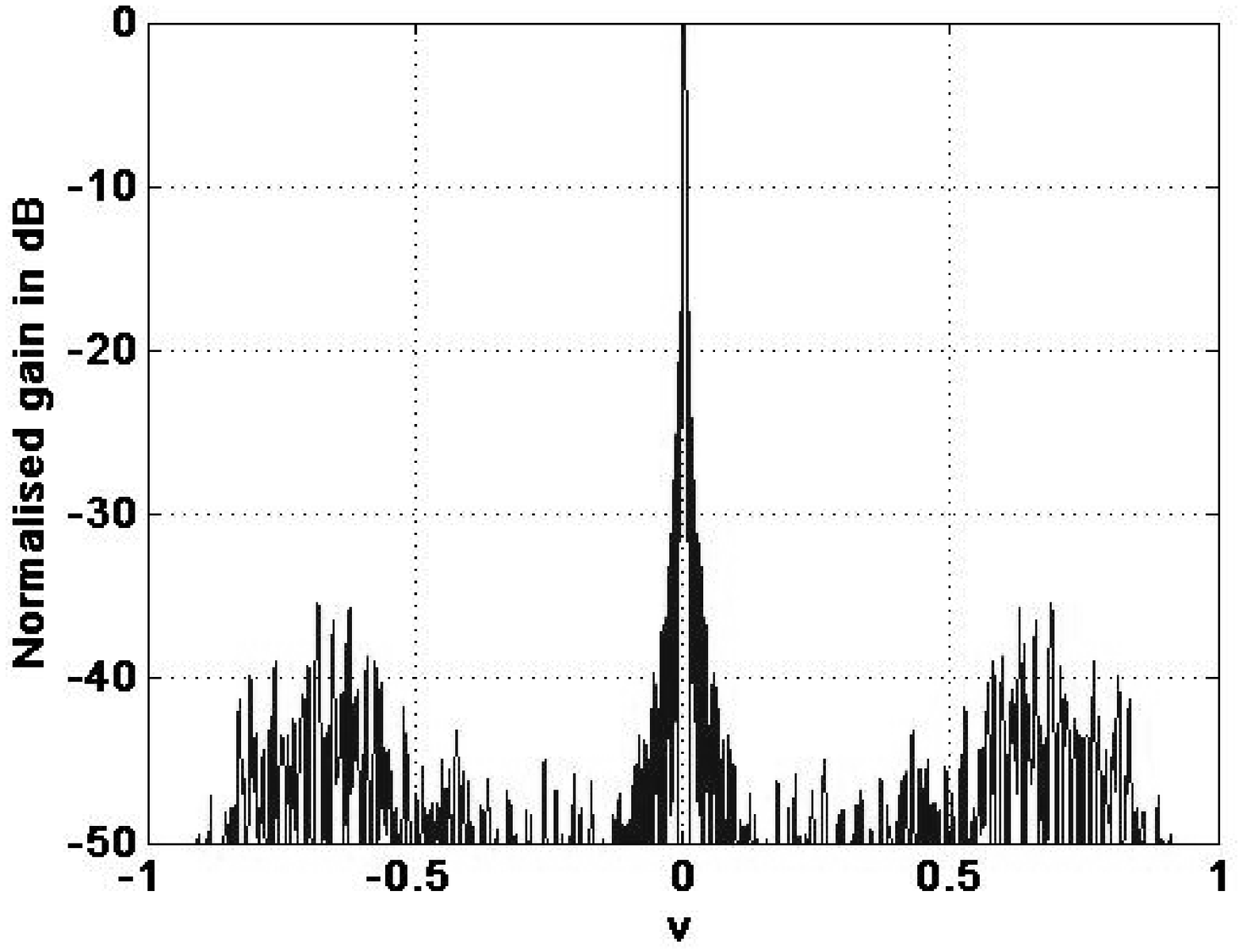}}
	\par\end{centering}
	
	\begin{centering}
	\subfloat[\label{fig:VO:2stage-reg-irreg:300}]{\centering{}\includegraphics[width=\columnwidth]{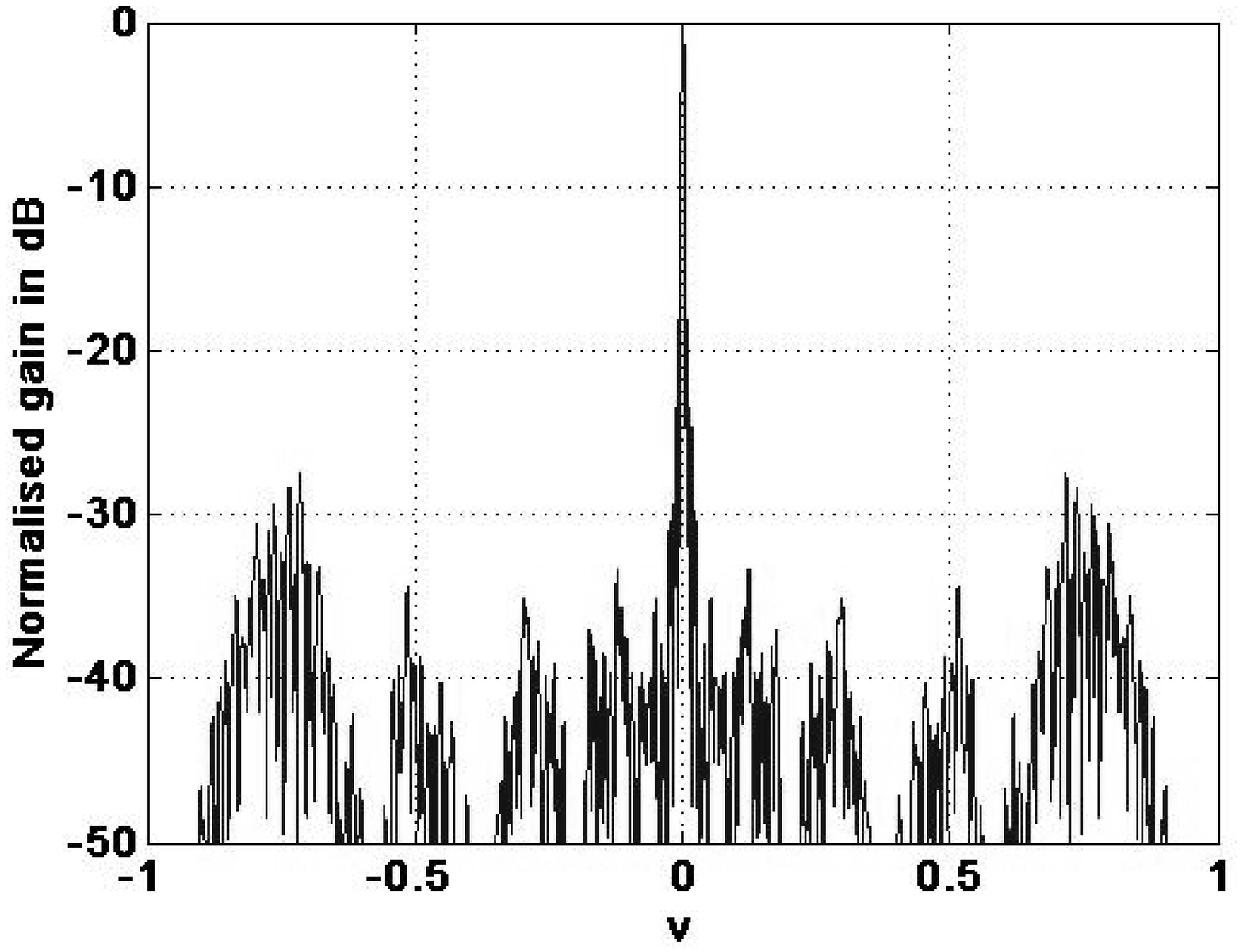}}
	\par\end{centering}
	
	\begin{centering}
	\subfloat[\label{fig:VO:2stage-irreg-irreg:300}]{\centering{}\includegraphics[width=\columnwidth]{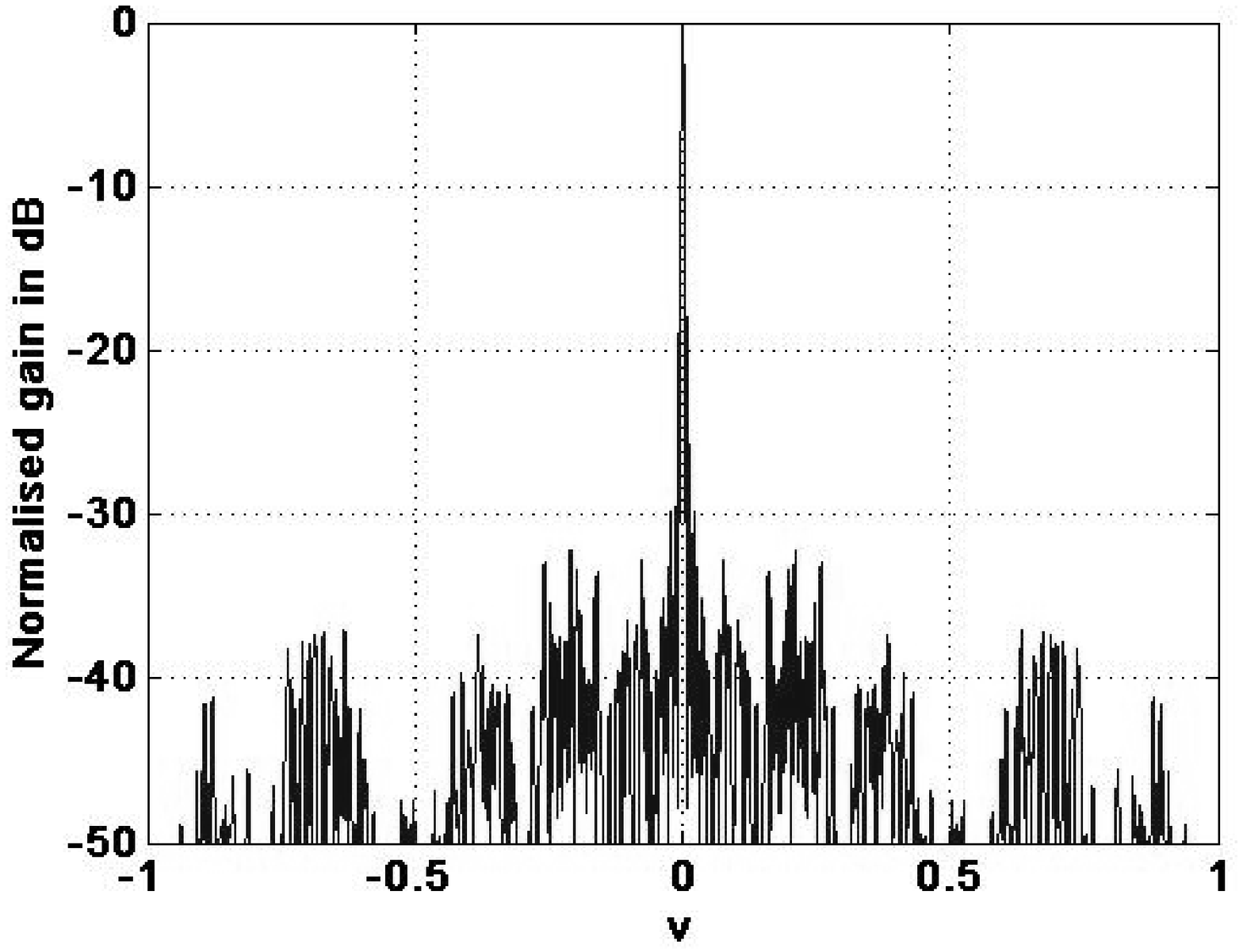}}
\par\end{centering}

	\caption{Broadside station beam at 300~MHz in v=0 plane for (a) single-stage irregular station layout, (b) two-stage regular-irregular station layout  and (c) two-stage irregular-irregular station layout. 
	\label{fig:VO:station-beam:300}
}
\end{figure}

The general trend in Figure~\ref{fig:aeff} is that the regular--irregular station exhibits the strongest departure from $\lambda^2$ at lower frequencies, and the two-stage irregular station the weakest. This is to be expected, as the minimum inter-element spacing of the regular--irregular layout is 1.3~m for every element. Meanwhile, the single-stage irregular station has some elements of larger minimum spacing, but most are still less than 1.4~m (see Figure~\ref{fig:1stageStats}). For the larger diameter two-stage irregular station, the irregular tile is larger, hence more of the elements remain sparse at lower frequencies.

The regular--irregular and single-stage irregular stations show a significant divergence in effective area at lower frequencies, despite having the same 189~m diameter. The greatest percentage difference in effective area occurs at $\approx$150~MHz,  where \AEff{} of the regular--irregular station is 74\% of the single-stage irregular station, equating to a similar loss in telescope sensitivity. An inter-element spacing greater than 1.3~m for the regular tiles is a potential solution, but the tile geometry puts limitations on such an increase because it restricts the randomisation of the station layout. A conclusive analysis of the potential loss of effective area due to regular tiles requires a full mutual coupling analysis and optimisation of the layouts, which is beyond the scope of this paper.

\subsection{Sidelobes}

As mentioned, over the full frequency range, the stations (for all layouts) operate as both dense and sparse array. The sidelobe characteristics, for the stations, will change as the array moves from dense to sparse region of operation. To observe these different characteristics, we analyse the sidelobes at two frequencies which are representative of the two regimes.

\subsubsection{Dense region (70~MHz)}

The primary sidelobe (sidelobe closest to main lobe) is independent of the intra-station layout. Figure~\ref{fig:station-beam} and Figure~\ref{fig:VO:station-beam} show a primary sidelobe level of approximately -18~dB for all three stations;
this level is consistent with that of a uniformly distributed circular aperture~\citep{Mai95}.

At the lower frequencies (eg. 70~MHz) the element separation is less than half wavelength. Thus, although the element (or tile) placement is random, the station aperture is `filled'. This `filled' aperture generates a station beam which contains an annulus of primary and secondary sidelobes. These sidelobes will not be smeared or suppressed by techniques such as station configuration rotation, which has been proposed for sparse arrays~\citep{CapWij06}. All three layouts exhibit these annuli. However, compared to the single-stage beamforming station, the sidelobes for the two-stage beamforming stations are spread over a wider area.

\subsubsection{Sparse region (300~MHz)}

At higher frequencies (eg. 300~MHz), for all stations shown in Figures.~\ref{fig:station-beam:300}--\ref{fig:VO:station-beam:300}, the sidelobes are suppressed due to the random placement of elements. However, the sidelobes for the single-stage beamforming station are concentrated towards the horizon, away from the main lobe. 

At the higher frequency of 300~MHz, the first-stage beamforming of tiles with relatively few elemental antenna inputs influences how the sidelobes are distributed as a function of scan angle and frequency. As Figure~\ref{fig:station-beam:300} shows, the irregular tiles smear out the strong, predictable sidelobes of the regular tiles. However, the general structure of the tile beam pattern is still visible in the station beam, especially for the regular tile. 

The level of the secondary station sidelobes visible in  Figure~\ref{fig:VO:station-beam:300} is affected by the first-stage beamforming. The maximum secondary sidelobe level increases from -35~dB for the single-stage station to -27~dB for the two-stage regular-irregular station.  In comparison, the classical 1/N result for distant sidelobes of a random distribution of 11\,200 isotropic antenna elements is approximately -40~dB. The irregular station does not reach this level because the randomisation is restricted to a relatively small range of element spacings as shown in Figure~\ref{fig:1stageStats}, resulting in the annulus of secondary sidelobes visible in Figure~\ref{fig:station-beam:300}.

The use of predictable sidelobes in calibration is under investigation~\citep{WijBre11, BraCap06}; if it is useful, then the regular tile beam structure may assist the calibration while maintaining a low sidelobe level. If more randomised sidelobe structure is beneficial, or even acceptable, significant gain in effective area are possible (see Figure~\ref{fig:aeff}).

\subsection{Station diameter and cost}

Although results will depend on the exact station layout, these examples illustrate the impact of the two-stage beamforming and inter-element spacing on effective area. These trends are important, as the current SKA system description proposes stations of  180\,m diameter and 11\,200 elements~\citep{Dewbij10}. The tight constraints on the minimum spacing due to the antenna footprint, and on the maximum spacing from the station diameter, limits the scope of layout optimization.

Given these constraints, then why not increase the diameter? An obvious feature of Figure~\ref{fig:aeff} is that the irregular--irregular station achieves a larger effective area at lower frequencies, for the same number of elements. Its larger diameter arises from the larger tile area required to achieve some degree of randomisation in the irregular tile, where the larger tiles simply do not fit in an irregular pattern within the 189~m diameter station.

The reasons not to place the same number of elements within a larger station diameter are two-fold: cost and calibratability. At higher frequencies, a larger station diameter reduces the station beam FoV with no commensurate increase in effective area. Therefore, fewer calibration sources are visible within the FoV, making the station calibration more difficult~\citep{WijBre11}.

The cost perspective is explored in \citet{ColHal12}. The larger physical area leads to increased costs related to on-site infrastructure. There are also increased signal processing costs to counteract the smaller station beam FoV.  The number of these beams required to cover a given area of sky (the processed FoV) scales as $\Dstmaths^2$. The digital station beamformer processing costs are approximately proportional to the number of inputs and the number of beams formed (see Appendix~\ref{app:beamforming-computational-cost}). The cost of data transmission from the station to the correlator and the correlator processing cost itself are also approximately linearly proportional to the number of station beams~\citep{ColHal12}. As an example, a 248~m diameter station requires 1.7 times more beams to achieve the same processed FoV as the 189~m diameter stations, resulting in roughly the same increase in signal processing costs.

\subsection{Future work}
Further to the illustrative examples in this paper, more complete optimisations will result in station layouts with slightly  improved performance. The representative layouts in this paper maintain a constant number of elements per station, resulting in a constant \AEff{} when the station is sparse at the higher frequencies. An alternative approach is change the number of elements or station diameter, to ensure that all layouts instead maintain a relatively constant \AEff{} at the lower frequencies. This will have cost implications, and the preferred optimisation approach depends on the SKA station requirements.

Regardless of the optimisation approach, there is scope for further characterisation of station performance, incorporating mutual coupling between the elements and the overall telescope response. For example, rotating the intra-station configuration between stations of the \low\ telescope helps to reduce sidelobes in the correlated beam~\citep{CapWij06}.  Similarly, rotating tiles within the station would smear out the station sidelobes arising from the regular tile layout. Increased randomization of elements or tiles in the station layout can reduce sidelobes, but requires more station area, resulting in a smaller station FoV. Down-weighting the edges of the aperture distribution through a spatial taper of the elements or tiles would reduce sidelobes~\citep{willey1962space}, again at the cost of increased station area.

\section{Conclusions}
We have quantified the first-order effect of two-stage beamforming on the station beam and effective area (\AEff) of the \low{}. We used both regular and irregular tiles within an irregular station layout and compared these to a single-stage irregular station. At higher frequencies the station \AEff\ is same irrespective of the inner station layout. At lower frequencies, regular tiles can significantly reduce the station \AEff\ compared to the other two configurations. Across all frequencies, the primary sidelobes are not affected by the station configuration, being largely a function of the station diameter. However, the secondary sidelobes are dependent on the input beam (tile or single element). While randomised two-stage configurations have the potential for considerable cost savings, the increase in far sidelobe levels by 10~dB or so will need to be assessed in the context of the SKA-low weighted science case.  For example, low sidelobe levels are required to characterise the ionosphere at low frequencies ~\citep{WijBre11} and enable precision imaging in e.g. the Epoch of Reionization domain. Conversely, time-domain studies at higher frequencies do not require precise secondary sidelobe control~\citep{StaHes11}. In combination with science priority setting, further station configuration studies which include, e.g., mutual coupling will be useful to refine studies of two-stage and other hierarchical beamforming architectures.

\appendix
\section*{Appendices}

\section{Tile beamforming cost reduction} \label{app_a}

\begin{figure}[h]
	\begin{centering}
	\subfloat[\label{fig:station-signal-path-all-digital}]{\centering{}\includegraphics[width=0.99\columnwidth]{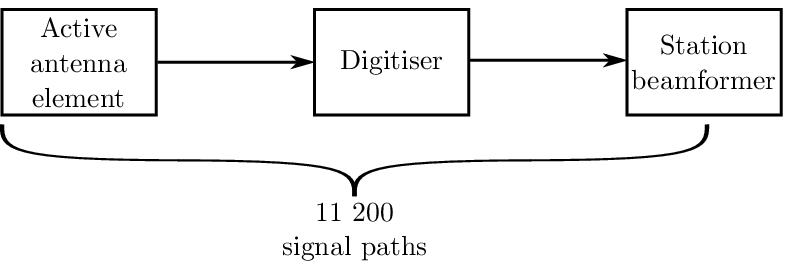}}
	\par\end{centering}
	
	\begin{centering}
	\subfloat[\label{fig:station-signal-path-RF-tile}]{\centering{}\includegraphics[width=0.99\columnwidth]{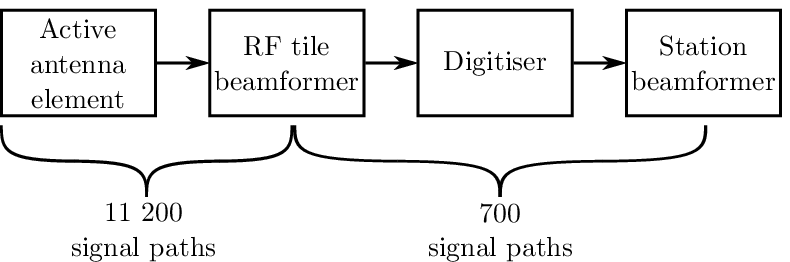}}
	\par\end{centering}
	
	\begin{centering}
	\subfloat[\label{fig:station-signal-path-dig-tile}]{\centering{}\includegraphics[width=0.99\columnwidth]{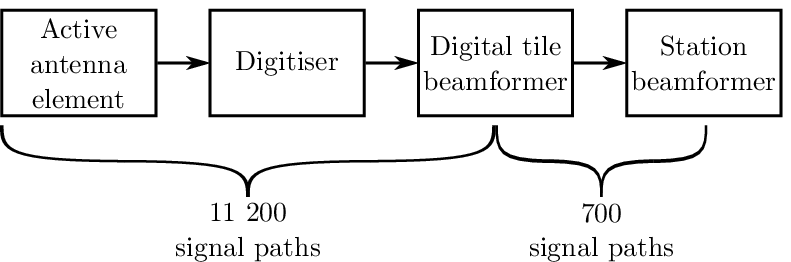}}
\par\end{centering}

\caption{Station signal path for (a) single-stage digital beamforming, (b) analogue (RF) tile beamforming and (c) digital tile beamforming. Arrows indicate signal transport (analogue or digital), but do not imply a particular physical location or signal transport technology. The number of independent signals is shown for a 11\,200 element station, with 16-element tiles for (b) and (c), where only a single tile beam is formed. \label{fig:station-signal-path}}
\end{figure}

We consider here how tile beamforming can reduce station hardware costs. The number of independent signals in a station can be determined by considering the elemental signal path. The signal path in the station begins at the active antenna element, culminating in one or more station beams as input to the correlator. In between, key actions are to digitise the signal, perform beamforming, and to transport the signal, in analogue or digital form, between the antenna element, digitiser and one or more beamformers. The intra-station signal transport and processing architecture defines the actions taken on the signal, and the order of these actions.

Figure~\ref{fig:station-signal-path} shows three such intra-station architectures, indicating the order of the actions and the number of independent signals at each point of the signal path. These are related to, but do not directly correspond with, the intra-station layouts in this paper. Figure~\ref{fig:station-signal-path-all-digital}, which shows a single-stage of digital beamforming, applies to the single-irregular layout. Figures~\ref{fig:station-signal-path-RF-tile}
and \ref{fig:station-signal-path-dig-tile} respectively show a first stage of analogue and digital tile beamforming, where the tile beamformer
is a signal aggregation point. Figures~\ref{fig:station-signal-path-RF-tile} and \ref{fig:station-signal-path-dig-tile} apply to both regular-irregular
and irregular-irregular layouts; we consider changes to the station beam pattern due to the different tile beamforming technology to be
second-order effects. As shown in Figure~\ref{fig:station-signal-path}, an intra-station architecture with tile beamforming reduces the number
of independent signals subsequent to the tile beamformer. Assuming only a single tile beam is formed, the reduction is the number of elements per tile, in our case 16.

We estimate the reduction in station hardware cost from tile beamforming by considering the number and cost of components in the elemental signal path. To a first-order approximation, the total cost for each these components in a station is linearly proportional to the number of independent signals. For the digital beamformers, the processing cost depends on the architecture, but is approximately linearly proportional to the  number of input signals and the number of output beams (see Appendix \ref{app:beamforming-computational-cost}). As mentioned, only a single output beam is assumed for the tile beamformer.  

The reduction in cost due to tile beamforming therefore depends on the cost of those components which are before and after the tile beamformer, and the fractional cost of these groupings. The total cost of components subsequent to the tile beamformer in Figures~\ref{fig:station-signal-path-RF-tile} and \ref{fig:station-signal-path-dig-tile}, such as signal transport to station beamformer and the station beamformer itself, decreases by a factor of 16. The cost of station hardware prior to the tile beamformer, such as that of the active antenna element, does not change with intra-station architecture. The total cost of digitisers in a station will be a factor a 16 larger when there is only digital beamforming (Figures~\ref{fig:station-signal-path-all-digital} and \ref{fig:station-signal-path-dig-tile}) than for analogue tile beamforming (Figure~\ref{fig:station-signal-path-RF-tile}).

An illustrative example using the fractional cost of the groups of components located before and after the tile beamformer shows that tile beamforming has the potential to significantly reduce station hardware costs. For example, consider a single-stage beamformed station. If the cost of the components prior to the tile beamformer represents half of the station hardware cost and those components after the tile beamformer represent the other half, then the two-stage beamforming cost is 53\,\% of the single-stage station hardware cost. Because an analogue tile beamformer (Figure~\ref{fig:station-signal-path-RF-tile}) aggregates the signals earlier on in the signal path than the digital tile beamformer (Figure~\ref{fig:station-signal-path-dig-tile}), more of the cost will occur after the tile beamformer, thereby further reducing the station hardware cost. 

\section{Beamforming computational cost \label{app:beamforming-computational-cost}}
Digital beamforming can be done in the time domain, or in the frequency domain on channelised signals. The computational cost of the frequency and time domain beamforming approaches is discussed in \citet{BarMil11} and \citet{KhlZar10}, where computation cost is expressed as a function of the number of output beams, input antennas and frequency channels,  with additional costs to implement a time delay (where necessary) and the FFT. Only the costs that scale with output beams and input antennas is relevant here; the other costs are specific to the signal processing architecture.

From \citet{BarMil11} and \citet{KhlZar10}, for a given architecture and number of channels, the frequency and time domain station beamformer processing load can be respectively simplified to 
\begin{equation}
P_{{\rm BF[\nu]}}\propto N_{\rm input}(K_{{\rm ch}}+K_{{\rm BF[\nu]}} N_{\rm beam})
\end{equation}
and
\begin{equation}
P_{{\rm BF[t]}}\propto N_{\rm beam}(K_{{\rm BF[t]}} N_{\rm input}+K_{{\rm ch}}),
\end{equation}
where $ N_{\rm input}$ is the number of elements or tiles being beamformed,
$ N_{\rm beam}$ is the number of station beams formed, the constant $K_{\rm ch}$
is the channelisation cost and $K_{{\rm BF[\nu]}}$
and $K_{{\rm BF[t]}}$ are the beamforming costs.

If the channelisation cost does not dominate (i.e. $ N_{\rm beam}\gg K_{{\rm ch}}/K_{{\rm BF[\nu]}}$
and $ N_{\rm input}\gg K_{{\rm ch}}/K_{{\rm BF[t]}}$), the cost scaling is
the same for both frequency and time domain beamforming.\label{ass:freq-time-BF-scaling}
The processing cost of the station beamformer is thus approximated
by 
\begin{equation}
P_{{\rm BF}}\propto N_{\rm input} N_{\rm beam}.\label{eq:cost-BF}
\end{equation}

\section*{Acknowledgments} 
We thank Dr. Adrian Sutinjo for the discussions surrounding this topic, and the reviewer for helpful suggestions in finalising our manuscript.

The International Centre for Radio Astronomy Research is a Joint Venture between Curtin University and the University of Western Australia, funded by the State Government of Western Australia and the Joint Venture partners. A. Jiwani is a recipient of a Curtin Strategic International Research Scholarship. T. M. Colegate is a recipient of an Australian Postgraduate Award and a Curtin Research Scholarship.

\noindent \bibliographystyle{apj}
\bibliography{bibfile}

\end{document}